# A new structural relaxation pathway of low-density amorphous ice

Jacob J. Shephard,[a] Stefan Klotz,[b] Martin Vickers[a] and Christoph G. Salzmann[*a]

[a] Department of Chemistry, University College London, 20 Gordon Street, London WC1H 0AJ, United Kingdom; E-mail: c.salzmann@ucl.ac.uk

[b] IMPMC, CNRS UMR7590, Université Pierre et Marie Curie, Paris, France.

**Abstract**

Low-density amorphous ice (LDA) is involved in critical cosmological processes and has gained prominence as one of the at least two distinct amorphous forms of ice. Despite these accolades, we still have an incomplete understanding of the structural diversity that is encompassed within the LDA state and the dynamic processes that take place upon heating LDA. Heating the high-pressure ice VIII phase at ambient pressure is a remarkable example of temperature-induced amorphisation yielding LDA. We investigate this process in detail using X-ray diffraction and Raman spectroscopy, and show that the LDA obtained from ice VIII is structurally different from the more 'traditional' states of LDA which are approached upon thermal annealing. This new structural relaxation pathway involves an increase of structural order on the intermediate range length scale. In contrast with other LDA materials the local structure is more ordered initially and becomes slightly more disordered upon annealing. We also show that the cascade of phase transitions upon heating ice VIII at ambient pressure includes the formation of ice IX which may be connected with the structural peculiarities of LDA from ice VIII. Overall, this study shows that LDA is a structurally more diverse material than previously appreciated.

**Introduction**

Low-density amorphous ice (LDA) is the most abundant form of ice in the Universe[1] and one of the at least two distinct polyamorphous forms of ice.[2] The kind of LDA that is thought to exist in space is called amorphous solid water (ASW) which forms by condensation of $H_2O$ from the gas phase onto cold surfaces.[3-5] In addition, it has been shown that LDA can be prepared by (i) hyper-quenching small water droplets at cooling rates greater than $10^7$ K s$^{-1}$ (HGW),[6-8] (ii) heating or decompression of high-density amorphous ice (HDA),[9,10] (iii) exposure of crystalline ice to UV radiation[11,12] or high-energy ion bombardment,[13] and (iv) heating the crystalline phases ice VII or ice VIII at ambient pressure.[14,15]

Originally, it has been argued that LDA from HDA is structurally different from ASW and HGW,[16-19] which were found to have similar structures.[16,20] Later, it was concluded that LDA from HDA, ASW and HGW are structurally identical.[21] Then subtle structural differences between LDAs prepared from HDA samples with different annealing histories were found and the two respective states were called LDA-I and LDA-II.[22] Consistent with an earlier study on LDA from HDA,[23] we have recently shown by using Raman spectroscopy that a continuous range of progressively more structurally relaxed states exists as LDAs from various different origins are heated towards the onset of crystallisation.[24] Interestingly, we could not confirm the existence of the distinct LDA-I and LDA-II states as defined in ref. 22 using Raman spectroscopy.

Apart from discussions on the structural differences of the various kinds of LDA there are also controversies with respect to the dynamic processes that take place upon heating LDA to its crystallisation temperature. Using calorimetry, a weak endothermic feature has been detected for annealed LDA samples just before the onset of crystallisation.[20,25-27] It has been suggested that the increase in heat capacity indicates a glass transition from the glassy state to the deeply supercooled low-density liquid.[20,28,29] However, the endothermic calorimetric feature is weak and other scenarios have been proposed as well such as the unfreezing of molecular reorientation dynamics.[24,30,31] It has been also been argued that the true glass to liquid transition of LDA is hidden above the crystallisation temperature and therefore difficult to observe experimentally.[32-34]

The formation of LDA by heating the crystalline high-pressure phase ice VIII at ambient pressure is a particularly interesting pathway for preparing LDA.[14] It has been suggested that this process takes place as the extrapolated melting line of ice VIII is crossed. For this reason, the ice VIII → LDA transition has been described as the 'symmetrical equivalent' in the phase diagram to the low-temperature pressure-induced amorphization of the 'ordinary' ice I$h$.[14] From computational work, the collapse of the ice VIII hydrogen-bonded network has been linked with highly non-linear behaviour of some vibrational modes of ice VIII upon decompression.[35,36] Ice VII, the hydrogen-disordered counterpart of ice VIII, has been shown to undergo a very similar transition to LDA upon heating at ambient pressure.[15] With densities of 1.490 and 1.483 g cm$^{-3}$ at ~90 K and ambient pressure,[37] ices VIII and VII are the two densest phases of ice that can be recovered at ambient pressure at liquid nitrogen temperature. Remarkably, all other crystalline high-pressure phases of ice transform to stacking disordered ice (ice I$sd$) directly upon heating without an intermediate stage of LDA.[38-41]

Here we investigate the formation of LDA upon heating ice VIII at ambient pressure in detail, its state of structural relaxation and the subsequent phase transitions using a combination of X-ray powder diffraction and Raman spectroscopy. The results are compared to the more 'traditional' states of LDA such as ASW and LDA from HDA, and discussed in the wider context of the complex structural and dynamic properties of the LDA material.

**Experimental**

Ice VIII samples were prepared by compressing 30 mg of ice I$h$ wrapped in indium foil in a hardened-steel pressure die to 2.5 GPa at 77 K using a 30-tonne hydraulic press. This was followed by isobaric heating to 260 K. Once the target temperature had been reached the samples were quenched with liquid nitrogen while keeping the pressure constant. Once the temperature reached 77 K the pressure was released and the samples were recovered at ambient pressure under liquid nitrogen. For further analysis, the ice was freed from the indium with a sharp blade.

An ice IX sample was prepared in a similar fashion by heating ice I$h$ at 0.3 GPa to 250 K. Unannealed high-density amorphous ice (uHDA) was obtained by compression of ice I$h$ at 77 K to 1.4 GPa according to Mishima's procedure.[9]

The ice samples were ground in liquid nitrogen and transferred into a purpose-built Kapton window sample holder which was mounted on a Stoe Stadi-P transmission diffractometer with Cu radiation at 40 kV, 30 mA, monochromated by a Ge 111 crystal to give K$\alpha_1$ only. Data were collected using a Mythen 1K linear detector and the temperature of the sample was controlled with an Oxford Instruments CryojetHT. The sample was oscillated in the beam to reduce texture effects from large grains and improve counting statistics. After quickly heating to a given temperature the samples were thermally equilibrated for 5 minutes before collecting diffraction patterns for about 10 minutes.

Small pieces of the samples were transferred into an Oxford Instruments Microstat[N] cryostat precooled to 80 K. The Raman spectra were recorded using a Renishaw Ramascope spectrometer equipped with a 632.8 nm He-Ne laser by using the continuous scanning mode and co-adding four spectra. For thermal annealing, the cryostat stage was heated at ~5 K min$^{-1}$ from 80 K to a given annealing temperature followed by cooling back to 80 K. The Raman shifts were calibrated using the sharp emission lines of a neon-discharge lamp.[42]

**Results and discussion**

The X-ray diffraction data recorded upon heating ice VIII at ambient pressure is shown in Fig. 1(a). The initial ice VIII material was phase-pure apart from a small contamination of ice I$h$ which resulted from vapour condensation during the sample transfer. The expected positions of the Bragg peaks of ices VIII and I$h$ are shown in Fig. 1(b). Consistent with earlier reports,[14, 38] the ice VIII sample started to transform to LDA upon heating at about 110 K as indicated by broad diffraction features centred at about 24 and 43 degrees. New crystalline features emerged at 125 K which can be assigned to ice I$c$ initially.[43] Upon further heating, the diffuse diffraction features of stacking disordered ice I (ice I*sd*) develop (*e.g.* left and right of the ice I*c* peak at about 24 degrees.).[44-46]

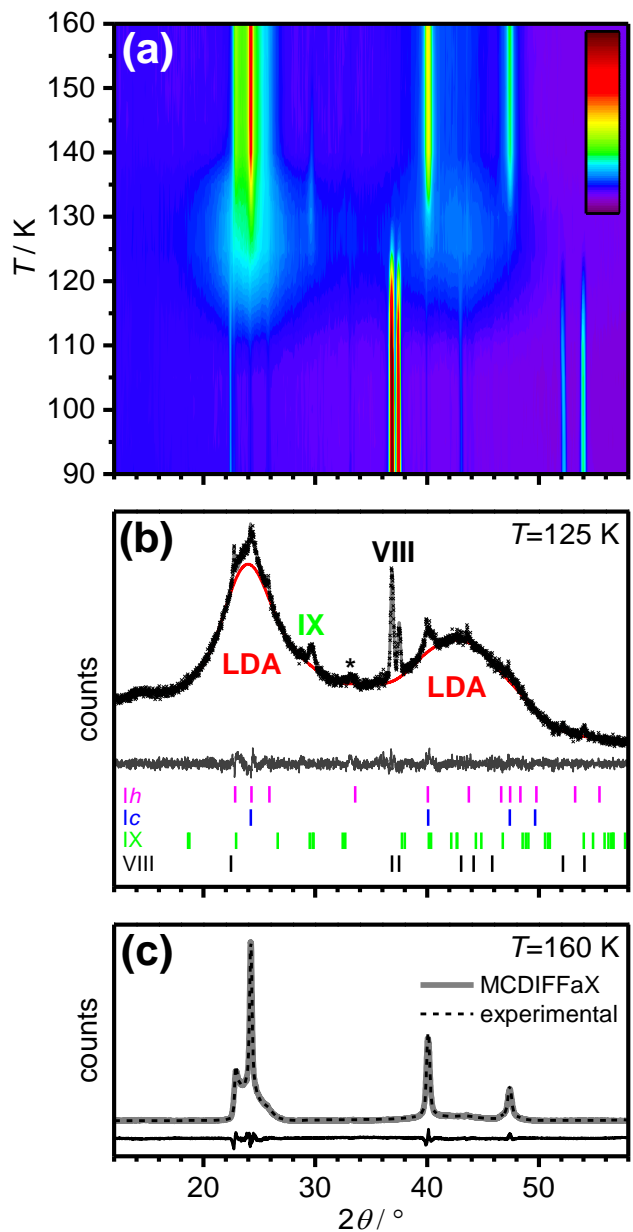

*Fig. 1 (a) Contour plot of the X-ray diffraction patterns recorded upon heating ice VIII at ambient pressure. To emphasise features with lower intensities the square roots of the intensities have been plotted. (b) Rietveld fit of the X-ray diffraction pattern at 125 K. The features of the amorphous LDA were fitted with a shifted Chebyshev polynomial. Tickmarks indicate the positions of the Bragg peaks of the various crystalline phases of ice. A feature resulting from the sample holder is marked with an asterisk. (c) Fit of the diffraction pattern of the stacking disordered ice I (ice Isd) at 160 K with MCDIFFaX.[46, 47]*

The new Bragg peak at 29.6 degrees ($d = 3.00$ Å) has been previously observed and been attributed to high-density amorphous ice (HDA).[48, 49] However, as pointed out in ref. 15 the sharpness of this peak is inconsistent with the expected broad diffraction features of HDA. Fig. 1(b) shows a Rietveld fit of the X-ray diffraction data at 125 K. The feature at 29.6 degrees is the most intense diffraction peak of IX which is the hydrogen-ordered counterpart of hydrogen-disordered ice III.[50] It is difficult to distinguish between ice III and ice IX on the basis of X-ray diffraction data. However, considering that the ice III to ice IX hydrogen ordering phase transition reaches completion above 165 K upon cooling at 0.23 GPa[51] we assume that ice IX forms at 125 K upon heating ice VIII at ambient pressure.

The relative amounts of the various phases observed upon heating can be calculated from the scale factors obtained from the Rietveld refinements.[52] The amorphous LDA cannot be included in the refinements and its contributions to the diffraction data were fitted as part of the background with shifted Chebyshev polynomials. Assuming that the amount of sample in the X-ray beam remains constant upon heating it is possible to estimate the weight percentages of LDA since these are the only unknown quantity. As mentioned earlier, the ice I which forms initially from LDA can be described with the structural model of ice I*c*.[43] Yet, at 135 K the features of stacking disorder develop very clearly and the ice I*c* structural model can no longer be used. Since there are now two non-crystalline components present in the sample it is not possible to estimate their weight percentages at 135 K. From Fig. 1(a) it can be seen that the LDA has disappeared completely by 140 K. Since there is now again only one non-crystalline component (ice I*sd*) the weight percentages of all components can be calculated again for temperatures greater than 140 K.

Fig. 2 shows the weight percentages determined in this fashion. Upon heating to 120 K about half of the ice VIII has transformed into LDA. The ice VIII content then decreases to about 10 w% at 125 K. This leads to the additional formation of LDA which reaches a maximal amount of 83 w%. This is consistent with earlier studies which have stated on a qualitative basis that it is not possible to obtain pure LDA by heating ice VIII at ambient pressure.[14, 48, 49] Alongside the formation of LDA small amounts of ice I and ice IX appear at 125 K. While the amount of LDA plateaus the weight percentages of ice IX and ice I in particular increase upon further heating to 130 K. At 140 K, the LDA has disappeared completely which produced large amounts of ice I*sd* but did not lead to an increase in the amount of ice IX. The ice IX impurity reaches a maximal amount of 5 w% between 130 and 140 K before it disappears completely at 160 K. The observed thermal behaviour of ice IX agrees well with a previous calorimetric study (*cf.* Fig. 2. in ref. 38). At this temperature, the sample finally consists entirely of ice I*sd*. Figure 1(c) shows that the diffraction data of ice I*sd* at 160 K can be fitted with our MCDIFFaX software[46, 47] which performs the refinement of cubic / hexagonal stacking probabilities with up to 2$^{nd}$ order memory effects ($\Phi_{ccc}$=0.7815, $\Phi_{hcc}$=0.3522, $\Phi_{chc}$=0.5098 and $\Phi_{hhc}$=0.6920). These stacking probabilities give an overall cubicity, *i.e.* percentage of cubic stacking, of 60.45% which lies below the currently highest reported value of 73.3%.[46]

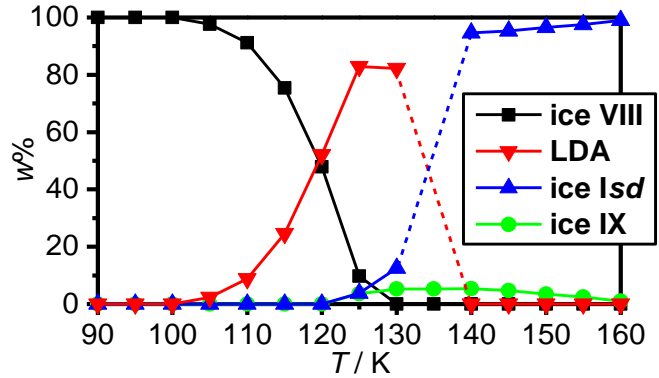

***Fig. 2*** *Weight percentages of the various phases of ice obtained upon heating ice VIII at ambient pressure. The dashed lines indicate that the weight percentages are unknown in this temperature range.*

Due to the complex cascade of phase transitions observed upon heating ice VIII at ambient pressure it is difficult to conclude with certainty on the exact pathways of the phase transformations. It is obvious that LDA forms directly from ice VIII and that LDA transforms to ice I$sd$. Yet, it is unclear if the ice IX forms from the last remaining ice VIII or from LDA.

A particularly interesting question is whether the LDA formed during the early stages of the ice VIII → LDA phase transition is structurally different from the LDA at higher temperatures. Fig. 3 shows the diffraction patterns of LDA from ice VIII, herein called LDA(ice VIII), at 115, 120 and 125 K. For this comparison, the diffraction patterns have been baseline corrected as indicated by the dashed lines and normalised with respect to the intensity of the first strong diffraction peak (FSDP) at about 22 degrees. The intensity of the diffraction feature at about 43 degrees decreases significantly upon increasing the temperature indicating progressive structural changes. For comparison, we also show the diffraction data of LDA obtained by heating uHDA, LDA(uHDA), at ambient pressure in Fig. 3. For LDA(uHDA), the relative intensities of the two diffraction features stay the same within the limits of the error of our measurement. Overall, it can be seen that the diffraction characteristics of LDA(ice VIII) approach those of LDA(uHDA) as the temperature increases but the two LDA materials have still not converged to the same structure at 125 K. Due to the presence of large amounts of ice I$sd$ in the LDA(ice VIII) material at 130 K it is difficult to normalise the diffraction data and it has been excluded from this comparison. It is noteworthy that in our previous Raman spectroscopic study we have detected a structural relaxation process in LDA(uHDA)[24] which is not seen here in the X-ray diffraction data indicating that Raman spectroscopy is more sensitive towards detecting structural relaxation in LDA compared to diffraction. Considering that the diffraction data of LDA(ice VIII) changes significantly upon heating suggests that this material undergoes substantial structural changes. Since our analysis relies on the normalisation of the diffraction data with respect to the FSDP it is unclear as to whether the structural changes in LDA(ice VIII) lead to a decrease of the intensity of the diffraction feature at 43 degrees, an increase of the intensity of the FSDP or a combination of changes in both features.

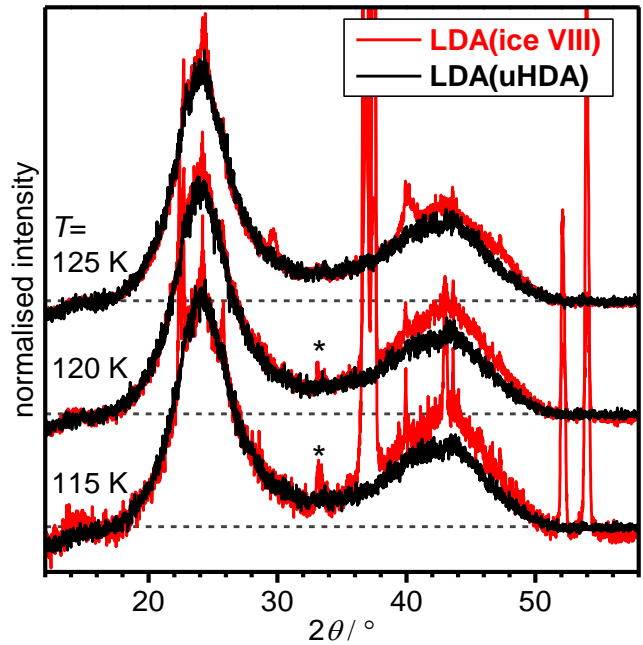

*Fig. 3* Comparison of the X-ray diffraction patterns of LDA from ice VIII with LDA from uHDA at 115, 120 and 125 K. The patterns have been baseline corrected as indicated by the dashed lines, normalised with respect to the intensities of the first strong diffraction peaks and shifted vertically for clarity. Features resulting from the sample holder are marked with asterisks.

Changes in the relative intensities of the diffraction features have been previously observed for ASW.[20] Upon annealing as-made ASW at 115 K the FSDP was found to sharpen and increase in intensity. Differences in the intensity and shape of the FSDP were also found for LDA-I and LDA-II[22] which we believe to be two states along the 'standard' relaxation pathway of LDA.[24] To further investigate the state of structural relaxation of LDA(ice VIII) we employ the same Raman spectroscopic methodology as previously used for ASW and LDAs from different states of HDA.[24]

The Raman spectrum of the ice VIII starting material in the coupled $\nu_{OH}$ stretching region is shown in Fig. 4 and it is in very good agreement with previous studies.[37, 53] To convert the ice VIII to LDA, the sample was heated in the cryostat while continuously monitoring its Raman spectrum. The phase transition to LDA was observed at ~130 K in this experiment and the sample was cooled back to 80 K in order to record its Raman spectrum. To study the effects of thermal annealing the LDA sample was then heated to increasingly higher annealing temperature always followed by immediate cooling back to 80 K after every annealing step to record the spectra. This temperature profile is needed to only observe the spectral changes due to structural relaxation of LDA and to exclude the effects of reversible thermal expansion.[24] A consequence of this temperature profile is, however, that the temperature scale is somewhat shifted with respect to the X-ray diffraction data presented earlier. Annealing at 155 K finally led to the crystallisation of LDA and the top spectrum in Fig. 4 is that of ice I*sd*.[54]

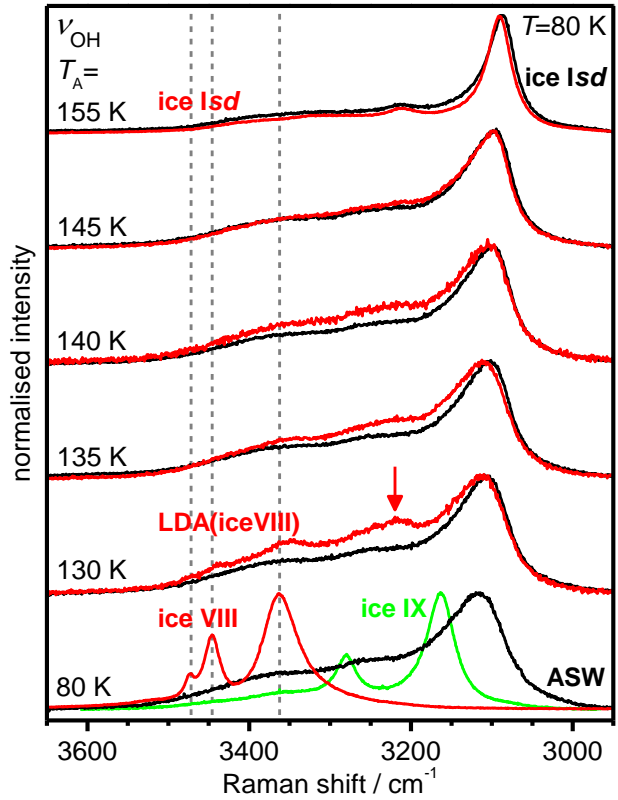

*Fig. 4* Raman spectra in the coupled $\nu_{OH}$ stretching region showing the irreversible phase transitions of ice VIII to LDA and ice Isd upon thermal annealing. The corresponding spectra of ASW from ref. 24 are shown for comparison as well as a spectrum of ice IX. All spectra have been recorded at 80 K after annealing at the indicated temperatures, normalized with respect to the intensity of the most intense feature and shifted vertically for clarity. A spectroscopic feature characteristic for LDA(ice VIII) is highlighted by the arrow. The dashed vertical lines indicate the peak positions of VIII.

In ref. 24 we have shown that the structural relaxation upon thermal annealing of ASW and LDAs from different states of HDA goes along with (i) a sharpening of the most intense feature of $\nu_{OH}$, (ii) a shift towards lower wavenumbers of this feature and (iii) decreases in the intensities of the shoulders on both the high and low wavenumber side of the most intense feature. From the spectra shown in Fig. 4 it can be seen that similar yet compared to ASW even more pronounced spectroscopic trends are found for the annealing of LDA(ice VIII). The spectroscopic differences between LDA(ice VIII) and ASW are greatest at 130 K. Upon thermal annealing, the spectra of the two kinds of LDA become increasingly more similar. Within the margins of error the $\nu_{OH}$ features of ASW were identical to those of LDA from various different HDAs provided that the same thermal annealing procedures are followed.[24] From the spectroscopic perspective, LDA(ice VIII) is therefore different compared to the 'traditional' forms of LDA.

The sharpening of the main feature of the coupled $\nu_{OH}$ mode upon thermal annealing has been interpreted in terms of an increase in structural order on an *intermediate* range length scale.[24] This argument was based on the fact that the O–H oscillators in ice couple with each other beyond the immediate local environment of a water molecule.[55] Along these lines, it can be concluded that LDA(ice VIII) becomes increasingly more structural ordered on an intermediate length scale upon thermal annealing. Considering the enormous density change during the ice VIII to LDA phase transition from 1.49 to 0.93 g cm$^{-3}$ we suggest that the build-up of stress and strain in the sample contribute to the structural disorder on a more extended length scale which can be reduced by thermal annealing.

The question now arises if LDA(ice VIII) immediately after its formation from ice VIII is simply a more unrelaxed state of LDA than the more 'traditional' LDAs or if it is an amorphous state that is unrelaxed in a somehow different fashion. Close inspection of the LDA(ice VIII) spectrum at 130 K shows a broad peak at about 3220 cm$^{-1}$ which is highlighted with a vertical arrow in Fig. 4. This feature is not thought to originate from any of the crystalline impurities that have been observed in X-ray diffraction. The ice VIII spectral features are located at considerably higher wavenumbers as indicated by the dashed vertical lines in Fig. 4. According to the earlier discussed X-ray diffraction data ice IX is expected to form together with the ice I$sd$ at higher temperature. But also, as shown in Fig. 4, the spectral features of ice IX would also not be able to produce the peak at 3220 cm$^{-1}$. It is therefore concluded that this spectral feature is characteristic for LDA(ice VIII). Since this feature cannot be seen in any of the ASW or LDA(HDA) spectra we conclude that the way LDA(ice VIII) is structurally unrelaxed is different compared to the 'traditional' states of LDA. Upon thermal annealing this characteristic feature disappears gradually.

The vibrational coupling of stretching vibrations in ice can be 'switched off' by adding small amounts of D$_2$O to the H$_2$O water from which the ice samples are prepared. This results in spatially well-separated O–D oscillators which can vibrate in a decoupled fashion. The spectral features in the decoupled $\mu_{OD}$ region are therefore thought to be only sensitive for the immediate local environment of the water molecules.

Fig. 5 shows the decoupled $\mu_{OD}$ region of LDA(ice VIII) containing 9 mol% HDO molecules. In line with the $\nu_{OH}$ spectra shown in Fig. 4 ice VIII components can be seen at about 2530 cm$^{-1}$ which decrease in intensity upon thermal annealing as the ice VIII weight percentages reduce. Increasing the annealing temperature from 130 to 135 and 145 K shifts the peak maximum of $\mu_{OD}$ feature of LDA(ice VIII) from 2436.2 to 2435.1 and 2433.9 cm$^{-1}$, respectively, while the half-widths *increase* slightly from 67.9 to 68.3 and 68.6 cm$^{-1}$. The small *increase* in half-width is remarkable since it is in contrast to what has been seen previously for ASW for which a *decrease* in the half-width of more than one wavenumber per 10 K was observed upon thermal annealing (*cf.* Fig. 3 in ref. 24). Similarly, for LDA(uHDA) the decoupled O–H feature became sharper upon thermal annealing.[56] Unlike in the cases of the 'traditional' LDAs the local structure of LDA(ice VIII) therefore seems to become slightly more *disordered* upon thermal annealing.

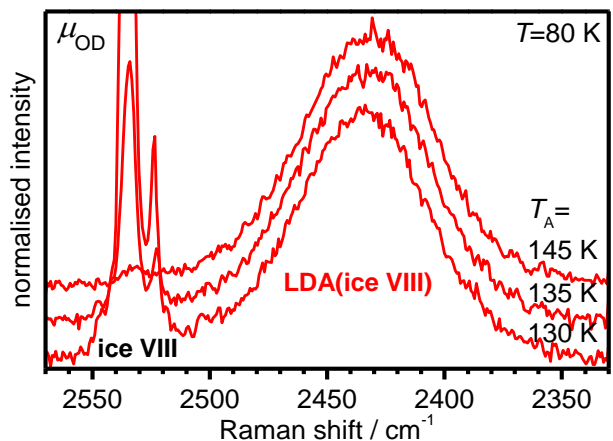

*Fig. 5* Raman spectra of the decoupled $\mu_{OD}$ stretching mode region at 80 K of $H_2O$ samples containing 9 mol% HDO. The spectra were recorded from LDA samples obtained from ice VIII at the indicated annealing temperatures, normalised to the maximum intensity of the LDA peaks and shifted vertically for clarity.

**Conclusions**

To summarize, our X-ray diffraction data indicates that LDA(ice VIII) undergoes pronounced structural changes upon thermal annealing. From the spectroscopic point of view the LDA(ice VIII) during the early stages of the ice VIII → LDA phase transition, is structurally more *disordered* on the intermediate range length scale and more *ordered* on the local length scale than more 'traditional' states of LDA such as ASW and the LDAs from HDA. The structural disorder on the intermediate length scale is attributed to stress and strain that result from the large density change during the ice VIII → LDA transition. The remarkable local order in LDA(ice VIII) is thought be a remnant of the crystalline order of ice VIII. It is not entirely clear if the ice IX impurity forms from the last remaining ice VIII or from LDA. Yet, it seems possible that it is the more ordered local structure of LDA(ice VIII) compared to the 'traditional' LDAs that opens up the pathway to the crystallisation to ice IX.

These findings illustrate that LDA is overall a more diverse material than previously appreciated – both as far as its structural complexity is concerned as well as its relaxation pathways. This study also shows once more that the structural relaxation of LDA is generally *slow* below the onset of crystallisation to ice I$sd$. The relaxation pathways of LDA(ice VIII) and of the 'traditional' LDAs approach each other upon thermal annealing. It is not entirely clear if they fully converge before crystallisation to ice I$sd$ sets in. Two non-converging relaxation pathways could not be reconciled with the postulated glass to liquid transition of LDA. At temperatures above such a transition the sample would have to be in equilibrium on the experimental timescale meaning that only one state can exist.

**Acknowledgements**

We thank the Royal Society for a University Research Fellowship (CGS, UF100144), the Leverhulme Trust for a Research Grant (RPG-2014-04) and J. K. Cockcroft for access to the Cryojet.